\documentclass[sigconf,screen]{acmart}

\AtBeginDocument{%
  \providecommand\BibTeX{{%
    \normalfont B\kern-0.5em{\scshape i\kern-0.25em b}\kern-0.8em\TeX}}}

\copyrightyear{2023} \acmYear{2023} \setcopyright{acmlicensed}\acmConference[CIKM '23]{Proceedings of the 32nd ACM International Conference on Information and Knowledge Management}{October 21--25, 2023}{Birmingham, United Kingdom} \acmBooktitle{Proceedings of the 32nd ACM International Conference on Information and Knowledge Management (CIKM '23), October 21--25, 2023, Birmingham, United Kingdom} \acmPrice{15.00} \acmDOI{10.1145/3583780.3614841} \acmISBN{979-8-4007-0124-5/23/10}

\usepackage{siunitx}
\usepackage{adjustbox}
\usepackage{caption}
\usepackage{subcaption}
\usepackage{todonotes}
\setuptodonotes{inline}
\newcommand{\acite}[1]{{\protect\NoHyper\citeauthor{#1}\protect\endNoHyper}~\cite{#1}}
\newcommand{\eg}{\textit{e.g.,}}
\newcommand{\ie}{\textit{i.e.,}}

\begin{document}

\title{Designing and Evaluating Presentation Strategies for Fact-Checked Content}


\author{Danula Hettiachchi}
\orcid{0000-0003-3875-5727}
\affiliation{
\institution{RMIT University}
  \city{Melbourne}
  \country{Australia}
}
\email{danula.hettiachchi@rmit.edu.au}

\author{Kaixin Ji}
\orcid{0000-0002-4679-4526}
\affiliation{
\institution{RMIT University}
  \city{Melbourne}
  \country{Australia}
}
\email{kaixin.ji@student.rmit.edu.au}

\author{Jenny Kennedy}
\orcid{0000-0001-9489-5471}
\affiliation{
\institution{RMIT University}
  \city{Melbourne}
  \country{Australia}
}
\email{jenny.kennedy@rmit.edu.au}

\author{Anthony McCosker}
\orcid{0000-0003-0666-3262}
\affiliation{
\institution{Swinburne University of Technology}
  \city{Melbourne}
  \country{Australia}
}
\email{amccosker@swin.edu.au}

\author{Flora D. Salim}
\orcid{0000-0002-1237-1664}
\affiliation{%
  \institution{The University of New South Wales} 
  \city{Sydney}
  \country{Australia}
}
\email{flora.salim@unsw.edu.au}

\author{Mark Sanderson}
\orcid{0000-0003-0487-9609}
\affiliation{
\institution{RMIT University}
  \city{Melbourne}
  \country{Australia}
}
\email{mark.sanderson@rmit.edu.au}

\author{Falk Scholer}
\orcid{0000-0001-9094-0810}
\affiliation{
\institution{RMIT University}
  \city{Melbourne}
  \country{Australia}
}
\email{falk.scholer@rmit.edu.au}

\author{Damiano Spina}
\orcid{0000-0001-9913-433X}
\affiliation{
\institution{RMIT University}
  \city{Melbourne}
  \country{Australia}
}
\email{damiano.spina@rmit.edu.au}

\renewcommand{\shortauthors}{Danula Hettiachchi~et~al.}
\begin{abstract}
With the rapid growth of online misinformation, it is crucial to have reliable fact-checking methods. Recent research on finding check-worthy claims and automated fact-checking have made significant advancements. However, limited guidance exists regarding the presentation of fact-checked content to effectively convey verified information to users. We address this research gap by exploring the critical design elements in fact-checking reports and investigating whether credibility and presentation-based design improvements can enhance users' ability to interpret the report accurately. We co-developed potential content presentation strategies through a workshop involving fact-checking professionals, communication experts, and researchers. The workshop examined the significance and utility of elements such as veracity indicators and explored the feasibility of incorporating interactive components for enhanced information disclosure. Building on the workshop outcomes, we conducted an online experiment involving 76 crowd workers to assess the efficacy of different design strategies. The results indicate that proposed strategies significantly improve users' ability to accurately interpret the verdict of fact-checking articles. Our findings underscore the critical role of effective presentation of fact reports in addressing the spread of misinformation. By adopting appropriate design enhancements, the effectiveness of fact-checking reports can be maximized, enabling users to make informed judgments.
\end{abstract}



\maketitle

\section{Introduction}

Fact-checking has become an integral part of online news and media journalism, and studies have shown that fact-checking can help curb the spread of online misinformation~\cite{Walter2020-ja,Nieminen2019-tw,Porter2021-sw,Carey2022-ob}. Duke Reporters' Lab, which has maintained a database of global fact-checking sites since 2014, reports that there were 378 active fact-checking projects operating in 105 countries in 2022~\cite{Stencel2022-sy}. Typically, these organizations review selected claims and compose fact-checking reports, primarily disseminated through their dedicated fact-checking website (\eg~ PolitiFact, Full Fact). This practice has led to a wide variety of fact-checking reports being made available to end-users who seek to fact-check information online.

However, there is limited guidance and consensus on how fact-checked content should be presented to effectively convey verified information to users. For example, veracity indicators (also truth scales, and visual rating scales) are common components that aim to communicate the final verdict in a summarized manner. While Duke Reporters' Lab states that about 80-90\% of the fact-checkers utilized veracity indicators and standardized labels to prominently convey their findings~\cite{Stencel2022-sy}, such indicators and the number of subcategories included can vary from one website to another. 

Effective presentation of fact-checking is an important research consideration which sits at the end of the fact-checking life-cycle \cite{spina2023human}. Prior works have focused on the problem of finding check-worthy claims~\cite{Hassan2017-gp}, automated fact-verification~\cite{Miranda2019-mb, Saeed2022-eg} or supporting fact-checking professionals by finding relevant sources, which represent the early and mid stages of the fact-checking process. 
Recent work by ~\acite{Pennycook2021-ka} notes that there is ``a large disconnect between what people believe and what they will share on social media, and this is largely driven by \textit{inattention} rather than by purposeful sharing of misinformation''. Therefore, presenting fact-checking information in a way that gets the attention of the user and accurately communicates the verdict is highly important. The entire effort that goes into rigorously verifying a claim and creating an insightful fact-checking report is futile, if the final verdict is misinterpreted due to presentation limitations. In this work, we set out to answer the following research questions:
\begin{itemize}
    \item \textbf{RQ1:} What are the important design elements in a fact-checking report?
    \item \textbf{RQ2:} To what extent can these elements improve users' ability to accurately interpret fact-checking reports?
\end{itemize}

To answer \textbf{RQ1}, we conducted a workshop to co-develop potential content presentation strategies. The workshops involved fact-checking professionals, communication experts, and researchers.
The participants explored the importance and efficacy of different elements (\eg~veracity indicators, explanations) and the feasibility of using interactive elements (components or conversation flows) that reveal additional information. 
To answer \textbf{RQ2}, drawing on the workshop outcomes, we then conducted an online experiment with 76 crowd workers to better understand the efficacy of different design strategies that aim to improve the credibility and the overall presentation of the fact-checking article. Our contributions are:

\begin{enumerate}
    \item From workshop findings, we establish the characteristics and features of effective presentation strategies for both screen and voice-based interfaces.
    \item We develop potential presentation enhancements, synthesize design recommendations, and introduce a crowdsourcing experimental setup to evaluate fact-checking presentation strategies.
    \item We demonstrate through our study that proposed presentation improvements can significantly improve users' ability to accurately interpret the verdict of the fact-checking articles.
\end{enumerate}

\section{Background and Related Work}

The spread of misinformation is a growing challenge that often raises concerns in domains like politics, public health and the environment~\cite{Carey2022-ob,Treen2020-ny}. False news spreads faster and more broadly compared to true news content~\cite{Vosoughi2018-oy}, and this problem has gained renewed attention with social media platforms enabling content sharing at scale.
Researchers have attempted to answer why people fall for and share misinformation. Attributes like political partisanship often emerge as a critical factor associated with overall belief, where people are more likely to believe in content concordant with their political partisanship~\cite{epstein2020will,Pennycook2021-ka}. Therefore, evidence from the literature suggests that research on misinformation-related topics should also consider the potential impact of user attributes.

While misinformation is predominantly shared and consumed on screen-based interfaces, with the growing availability of devices equipped with digital voice assistants, people consume potential misinformation through voice interaction~\cite{Ferrand2020-yr} and other audio-based channels such as podcasts~\cite{Caramancion2022-pl}.

\subsection{Fact-Checking Reports}

One fundamental approach to reduce online misinformation is fact-checking or verifying content or claims. Research provides overwhelming evidence that fact-checking has an overall positive influence in preventing misinformation online~\cite{Walter2020-ja,Nieminen2019-tw,Porter2021-sw,Carey2022-ob,ipie2023}. Fact-checking is often conducted by media organizations (\eg~AFP Fact Check, BBC Reality Check) or other independent organizations (\eg~Africa Check, FactCheck.org).

Fact-checking organizations generally disseminate their verdicts through online articles, often appearing on their dedicated websites. A fact-checked article composed by a fact-checking organization can include a variety of components such as the original content (or a link), verdict statement, explanation, veracity indicators, sources used to arrive at the verdict, credentials of the fact-checking organization, engagement options (\ie~commenting, voting, and sharing), and a summary snippet~\cite{Pal2019-ly}. In addition, the websites often provide details about the editorial committee or fact-checkers, fact-checking mechanism, and the veracity indicator or classification scale. However, there is limited consistency in presentation across organizations. As reported by \acite{Pal2019-ly}, out of 22 fact-checking websites examined, only the original content and explanation were available across all 22 websites. Other elements -- sources used to verify (in 21 websites), social media sharing option (in 21), veracity indicator (in 18), details about the fact-checking mechanism (in 17), and comment option (in 8) -- were not always available.

Veracity indicators (also truth scales, and visual rating scales) are graphical elements that aim to communicate the final verdict in a summarized manner (\eg~false, mostly true)~\cite{Amazeen2018-vx}. For example, PolitiFact includes a scale known as `Truth-O-Meter', which includes six ratings from True, Mostly true, Half true, Mostly false, False, to Pants on fire. While veracity indicators are widely used in popular fact-checking websites (\eg~PolitiFact, Snopes, RMIT ABC Fact Check), 
there is limited consensus on their effectiveness. \acite{Amazeen2018-vx} report a significant increase in fact-checking effectiveness when adding a veracity indicator to a contextual correction for non-political content. On the contrary, in a meta-analysis that includes 30 studies, \acite{Walter2020-ja} reports that fact-checking presentations that included visual accuracy cues produced significantly weaker effects compared to presentations that did not include an indicator.

Users' perception of the organization that conducts fact-checking can induce additional biases and impact the effectiveness of fact-checking~\cite{Brandtzaeg2017-yo}. Organizations can enhance their credibility by being part of global networks such as the International Fact-Checking Network (IFCN). 
However, research shows that users often pay little attention to cues about the source of information. For instance, including a log banner that highlights the publisher does not lead to a reduction in the susceptibility to misinformation~\cite{Dias2020-mp}.

When presenting fact-checking reports to users, using different channels in addition to regular fact-checking articles can be helpful. A study by~\acite{Young2018-ql} reports that both humorous and non-humorous short videos work better than text-based fact-checks due to increased clarity and attention. Chatbots are also shown to be an effective way to deliver fact-checking reports. Through a case study of health information, \acite{Zhao2023-pd} show that chatbots increase users' perceived ease of use, although they do not have any significant impact on users' fact-checking intentions.

\subsection{Other Approaches to Fight Misinformation}

Lack of scalability is a key limitation of fact-checking by professionals as it requires significant time and effort to investigate and verify a particular claim~\cite{Nieminen2019-tw, Haque2020-an}. Research on automated approaches attempts to address these challenges by complementing existing fact-checking infrastructure~\cite{das2023state, You2019-gp}.
These approaches include automated fact-checking systems that aim to predict a verdict using existing online sources~\cite{Miranda2019-mb,Trokhymovych2021-xt}, using crowd judgments to assess the veracity of statements~\cite{roitero2020can,Soprano2021-vn, Saeed2022-eg}, automated methods that can assist fact-checking teams by finding check-worthy claims~\cite{Hassan2017-gp, Smeros2021-je}, and improving search results for fact-checking~\cite{Yasser2018-bq}. Recent research has also explored ways of making automated fact-checking more interpretable and explainable~\cite{Zhang2021-tq}. While automated approaches directly support critical steps of the fact-checking pipeline, ultimate fact-checking decisions need to be published through fact-checking reports or similar mediums. Therefore, the work presented in this paper still complements research on automated approaches.

While fact-checking is vital to counter the spread of misinformation, researchers have also examined complementary approaches~\cite{Pennycook2021-nq, Jahanbakhsh2021-qf}.
For example, on social media, lightweight interventions at the time of posting, such as asking users to assess content accuracy and provide their reasoning, can deter users from posting misinformation~\cite{Jahanbakhsh2021-qf}. Similarly, encouraging people to think about accuracy can help reduce the sharing of misinformation. For instance, reporting on a Twitter study, \acite{Pennycook2021-nq} show that sending users private messages and asking them to rate the accuracy of a single non-political headline improves the quality of the news that they subsequently share. However, these approaches have not been adopted and implemented widely, and therefore, fact checking remains integral to curbing online misinformation.

\section{Exploring Presentation Strategies for Screen and Voice Interfaces}

Guided by the lack of consistency in the presentation of verified content and limited research on the effective presentation of fact-checking reports, we conducted a hands-on design workshop to develop a set of fact-checked content presentation strategies for screen interfaces. Through the workshop, we also aimed to understand the process of delivering fact-checked content through voice-only interaction and develop relevant basic presentation strategies. The 4-hour workshop involved 10 participants representing fact-checking professionals, communication experts and researchers. Participants were selected based on their expertise and availability after calling for an expression of interest. Participants received an AU\$60 gift voucher.  

The workshop included five hands-on activities and discussions. After a brief introduction session, in the first activity, participants were asked to identify and list different fact-checking elements by placing a sticky note on eight example fact-checking reports displayed on a wall. The reports were extracted from eight different global fact-checking websites such as AFP Fact Check and Full Fact. Participants were then allocated to three groups, ensuring diversity in expertise. In the second activity, as a group, participants discussed the elements they noted in activity one. They identified important elements for fact-checking presentations by ranking all the identified elements according to their importance. Participants shared and discussed their findings with other groups and facilitators in a debriefing session.

In the third activity, participants were asked to develop a fact-checking presentation strategy for screens. A presentation strategy may consist of elements identified in the previous activity or adapted versions of those elements. Participants were given Lego blocks and instructed to use different colors to represent the element type, vertical height to represent the level of expected attention/intensity, and the occupied surface area of the block to represent the space requirement on the website.

To foster in-depth discussions on how fact-checking presentations could impact different individuals, we developed six personas that were used in the next activities. Informed by the Australian Digital Inclusion Index~\cite{Thomas2021-dii,Thomas2023-qa}, each persona included the four characteristics in addition to basic demographic information. (i) Technology Attitude: including notions of control, enthusiasm, learning, and confidence toward technology; (ii) Basic Technology skills: including basic operational functions, such as connecting to the internet, downloading and opening files, using software, deciding what to share, how and who with, managing and monitoring contacts, and communicating with others; (iii) Advanced Technology skills: including information searching and navigating, verifying trustworthy information, and managing third-party data collection, adjusting privacy settings, determining what is safe to download, customizing devices and connections; and (iv) English Proficiency: the ability to read and write in English fluently.

In the fourth activity, each group received the six personas and was asked to pick two personas with contrasting characteristics considering their interest (or willingness) in consuming fact-checking content. Participants then created a scenario using the persona and a topic from the fact-checking reports shown in the first activity. It could include details such as news consumption channel, motivation to use the fact-checking website, etc. Participants could also add more details to the persona (e.g., demographics, familiarity with the topic, willingness to verify information). Afterwards, participants modified the fact-checking presentation they developed in the third activity for each persona/scenario and shared their outcomes with other groups and facilitators.

During the last activity, participants developed a fact-checking presentation strategy for voice-interaction. A presentation strategy consisted of elements identified before or adapted versions of those elements. Similar to the third activity, participants used Lego blocks where color reprinted the element type, vertical height represented the level of expected attention/intensity, and the order of elements (top to bottom) indicated the turns in conversation.

The workshop activities were approved by RMIT University’s Ethics Board (ID~24942), and all workshop activities and group discussions were recorded and transcribed after the workshop. For our analysis, we mainly focus on the debriefing sessions where participants shared and discussed their thoughts with facilitators and other groups after each activity. We apply the general inductive approach to data analysis as defined by \acite{Thomas2006-oe}. Two of the authors independently coded the transcripts, and then met with a third author to discuss and finalize the themes.

\subsection{Findings of a Collaborative Workshop}

\begin{figure}[htb]
    \centering
    \includegraphics[width=\columnwidth]{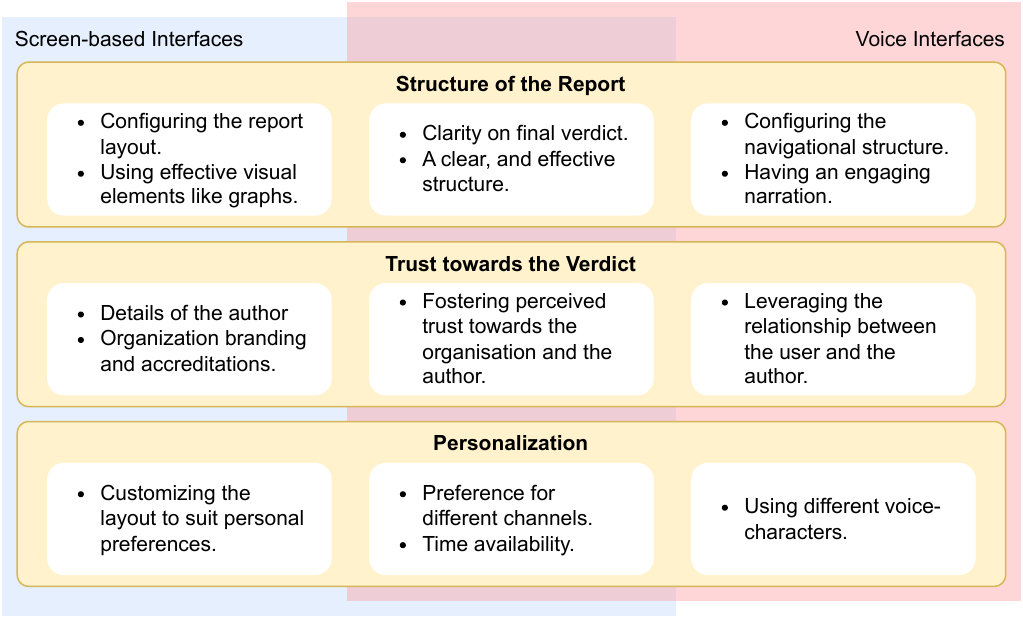}
    \caption{An overview of workshop outcomes.}
    \label{fig:workshop-outcomes}
\end{figure}

Figure~~\ref{fig:workshop-outcomes} summarizes the answers to \textbf{RQ1:} What are the important design elements in a fact-checking report? We identified three main themes concerning the structure of the report, trust towards the verdict, and effective personalized presentation. Under each section, we discuss design considerations for both screen-based and audio-based presentations.

\subsubsection{Structure of the Report}

Organizing the fact-checking report content into a clear and effective structure was perceived as an important design requirement in both screen and audio interactions. 

\begin{quote}
    ``We thought it's good to keep it simple for people and not overload with information. It's very simple [in terms of] explaining what's going on. Simple claim, verification and the explanation, data visualization.''~[G2]
\end{quote}

\begin{quote}
``These are not academic pieces. These are not designed for people with PhDs, especially with the elections and all, it's designed to inform you about what is going on in the world. We wanted to make it as simple as possible for people.''~[G2]
\end{quote}

Participants stressed the importance of \textit{providing clarity on the final verdict}, which can be achieved by properly wording the title of the fact-checking article and including explicit veracity indicators. In audio, the verdict could be stated prominently at the beginning of the conversation. In addition, considering the varied attention and focus of different groups and time-poor online users, participants suggested that having an \textit{explicit and informative summary} of the report could be useful. The summary should succinctly explain the claim, including the context of who, what, and why.

\begin{quote}
    ``We were sort of arguing that within the title, it's really important to summarize the claim and provide an indication of the overall verdict. True or false, kind of thing. 
    So, for example, we were looking at the Ivermectin one, like clearly identifying that is a false claim, in the title, so it's not necessarily hooking reason, sort of giving them that informative summary.''~[G1]
\end{quote}

\begin{quote}
    ``That was more like either an interview or some direct, um, direct info in that sense and then a bit more info. And then the full verdict at the end, reiterating the closing hook of the story and then again ending with the brand.''~[G3]
\end{quote}

In regular screen-based presentations, participants suggested including images, videos, or screenshots related to the claim to enhance understanding and engagement. Data visualization techniques were recommended to present complex information in a visually appealing and accessible manner.

\begin{quote}
    ``Also, an image or video of the claim. So a screenshot of you know what? Something stupid someone sent on Facebook or a video of the person saying it, if that's there''~[G2]
\end{quote}

The audio-based fact-checking presentation can take multiple forms. Mainly, the users could listen to a fact-checking report presented in a podcast-like audio clip, or they could consume fact-checking reports through a turn-based interaction with a voice assistant. In the remainder of the paper, for simplicity, we refer to these as podcasts and voice assistants.

In particular, when presenting fact-checking reports through voice assistants, the \textit{additional effort required to navigate the report} was a key concern. A user could initiate the fact-checking conversation with a simple prompt, but then they will need multiple follow-up prompts to know about the origin of the claim, the sources used for fact-checking etc. There is also the risk of users missing out on critical information.

\begin{quote}
    ``Imagine you are an average punter, who's like, you know, just want to fact-check, going on a game of 20 questions, where you're sort of like jumping all around the place. Yeah. I don't know. I just think it requires a lot of effort.''~[G1]
\end{quote}

\begin{quote}
    ``It's just a labyrinth. It's all interactive, but within the defined spaces you are allowed to go.''~[G2] 
\end{quote}

\begin{quote}
    `` You sort of like you've asked about something. We'll tell you kind of a summary of what we think this is. And then here are the options for more information because there would be a lot of different ways that you can answer.''~[G2]
\end{quote}

Diverging from the question-answering structure, using a \textit{narrative structure can engage users} in voice-only fact-checking. 
Participants mentioned the feeling of intimacy in different interaction forms. 

\begin{quote}
    ``Yeah, I think audio is also referenced as, like, the intimate, like the radio, the longevity of radio. The whole idea is, it's an intimate communication form, which is perhaps why we're leaning more towards a narrative.''~[G1] 
\end{quote}

\begin{quote}
    ``Potentially with audio, you can start off with someone saying like, I used to believe that climate change didn't exist and here's how I fell into that. I believe you can build your narrative so it allows you to maybe have some more [sympathy].''~[G1] 
\end{quote}

The length of the fact-checking report was another design consideration for podcast-like formats. Participants suggested generating both shorter versions, as well as longer informative versions that provide all information typically written in a fact-checking report.

\begin{quote}
    ``There's a five-minute version, and then there are longer versions. I think that's really important.''~[G3]
\end{quote}

\noindent Participants also mentioned that is useful to start the audio with ``a hint'' of the claim.

\subsubsection{Trust towards the Verdict}

Fostering trust towards the verdicts presented in fact reports is highly important. Participants noted that the trust towards the verdict presented in a fact-checking report could hinge on the users' perceived trust towards the fact-checking organization and the specific author/editor who investigated the claim and composed the fact-checking report. Participants argued that the \textit{credibility of the organization} plays a major role and noted that organizations could use clear branding, such as including the logo and having distinct color schemes and fonts. Such a strong `visual framework'~\cite{Tidwell2011-gm} will also make the report easier to use and navigate. In addition, accreditations, certifications, and other affiliations with reputed organizations could be used in the report.

\begin{quote}
    ``Our top element was the branding of the organization, so that we know that there's some authority behind what we're about to read.''~[G2]
\end{quote}

\begin{quote}
    ``Also clear affiliation and the site brand. So perhaps with some sort of clear branding, the logo and the coloring. So you would know that it's what it is, and hopefully, that's something recognizable as well, and then the certification.''~[G3]
\end{quote}

In addition to the organization, participants thought that the \textit{details of the author} could elevate trust. While providing information about the editorial committee or fact-checkers is a common practice in current fact-checking websites~\cite{Pal2019-ly}, individual reports are not always directly associated with specific authors. Participants noted that the relationship between the user and the author could play a prominent role in audio formats like podcasts.

\begin{quote}
    ``You do build a rapport with your podcasts, [and] the host that you don't have. As you said, like the author, I have an article because you're hearing that guy's voice over and over again.''~[G1]
\end{quote}

Some discussions highlighted using sympathy to establish trust with the system. Similar to a daily conversation, participants expected sympathy can help users better understand content. Participants pointed out that voice identification establishes sympathy/trust in real-world social interaction; this can be extended into voice assistance by providing options for different audio characters.

\begin{quote}
    ``So there is that narrative that human connection with some of I think it might have been. [The persona] said he really trusted people's voices. He trusted his postman. He trusted people rather than just unidentified technology. So I think the audio does have that voice, and it's probably a bit different if it's Alexa, compared to if it's a recorded real person as well about that trust element.''~[G3] 
\end{quote}

\subsubsection{Personalization}

Activities that utilized the personas led to insightful discussions around the personalization of fact-checked content. For users with different reading preferences, or time availability, personalizing the presentation styles can help consume the information more effectively. For instance, we can \textit{align the presentation format with users' existing preferences}, such as mimicking newspaper formats for older users or incorporating scrolling features for younger, tech-savvy users.

\begin{quote}
    ``It has a direct personal link, so it can be similar to the newspaper format where he can link it to a particular writer.''~[G3]
\end{quote}

\begin{quote}
    ``He's a young adult, very tech savvy, very quick, so he doesn't like to read long amounts of info, just wants the idea\ldots We decided to keep the subheadings and built them a little bit up. So he could just sort of quickly scroll through and see that.''~[G3]
\end{quote}

Simplifying the presentation of information, using concise explanations, and incorporating visual elements, as discussed above were also seen as beneficial for users with limited patience or tech proficiency.

\begin{quote}
    ``Specially, for both personas, the teenager is not going to go through it. I had trouble figuring out the article, so for the mom, it could be the same. It is very dense text. [We want] to use more images.''~[G2]
\end{quote}

In audio, preferred and familiar \textit{voice characters} can improve comprehension and engagement.

\begin{quote}
    ``You could have different characters, that are sympathetic, and more understanding.''~[G3]
\end{quote}

Participants also briefly mentioned the need for fact-checking reports to \textit{present information in an unbiased or impartial manner}. Excessive personalization could create a risk of additional biases. For instance, a concise report could lose context and therefore, the designers and developers should carefully assess and validate the effectiveness of any personalization propositions. 

\begin{figure*}[t]
    \centering
    \includegraphics[width=.8\textwidth]{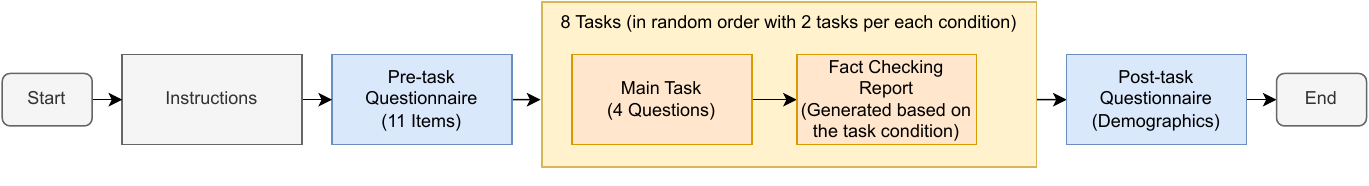}
    \caption{The sequence of questionnaires, tasks and surveys participants conducted during the online study.}
    \label{fig:study-flow}
\end{figure*}

\section{Crowdsourcing Experiments to Evaluate Designs}

Based on workshop outcomes and prior work on misinformation and fact-checking, we identified six key design elements of fact-checking reports and designed an online user experiment to investigate the impact of including such credibility and presentation elements in reports.\footnote{Code and data are available at \url{https://github.com/ADMSCentre/factchecking-presentations}.} By adding `Credibility' elements such as accreditation, additional author details, and sources, we aim to improve user trust towards the fact-checking reports. With `Presentation' elements, we aim to make it easy for users to navigate the report and find the relevant information they seek. Table~\ref{tab:experiment-features} provides details of the design elements used in our experiments.   

While our collaborative workshops focused on both screen and voice interfaces, our first attempt at experimental implementation and crowdsourced evaluation presented in this paper focused on screen-based interfaces. Currently, screen-based implementations (\eg~fact-checking websites) are predominantly used to disseminate fact-checking reports.

\begin{table}[tb]
    \centering
    \caption{The design elements investigated and details of how the fact-checking reports were rendered in the experiments.}
    \label{tab:experiment-features}
    \adjustbox{max height=0.17\textheight}{
    \begin{tabular}{lp{1.5cm}p{5.5cm}}
    \toprule
         {\it Category} & {\it Feature} & {\it Details} \\
         \midrule
         Credibility & Accreditation & Including the details about affiliations of the fact-checking organization (\eg~part of IFCN network). Accreditation is not included in the baseline.\\
         & Author \newline Details & Including a profile photo and a brief paragraph about their experience/qualifications. In the baseline, only the author's name is included. \\
         & Sources & Listing all the relevant sources used to create the article. No source details are included in the baseline.\\
         \midrule
         Presentation & Veracity \newline Indicator & Including a graphical veracity indicator that communicates the final verdict. In baseline, final verdict is mentioned in the text.\\
         & Structure & Organizing the main text with sub-headings and paragraphs. In the baseline, sub-headings are not included.\\
         & Summary & Including a short summary on top of the article. The summary is not included in the baseline.\\
         \bottomrule    
    \end{tabular}
    }
\end{table}

In our study, we used eight fact-checking articles in total, with four from PolitiFact and four from RMIT ABC Fact Check. 
All fact-checking reports were based on statements related to the economy, environment, or demographics -- with no subjective/ideological statements or ones directed at others -- made by politicians of the national level from two major parties in the US or Australia.
We balanced the dataset in terms of the veracity and the political party of the speaker (\ie~Democrat vs. Republican for the US; Labor vs. Liberal for Australia).

\begin{figure}[tb]
     \centering
     \begin{subfigure}[b]{0.37\textwidth}
         \centering
         \includegraphics[width=\textwidth]{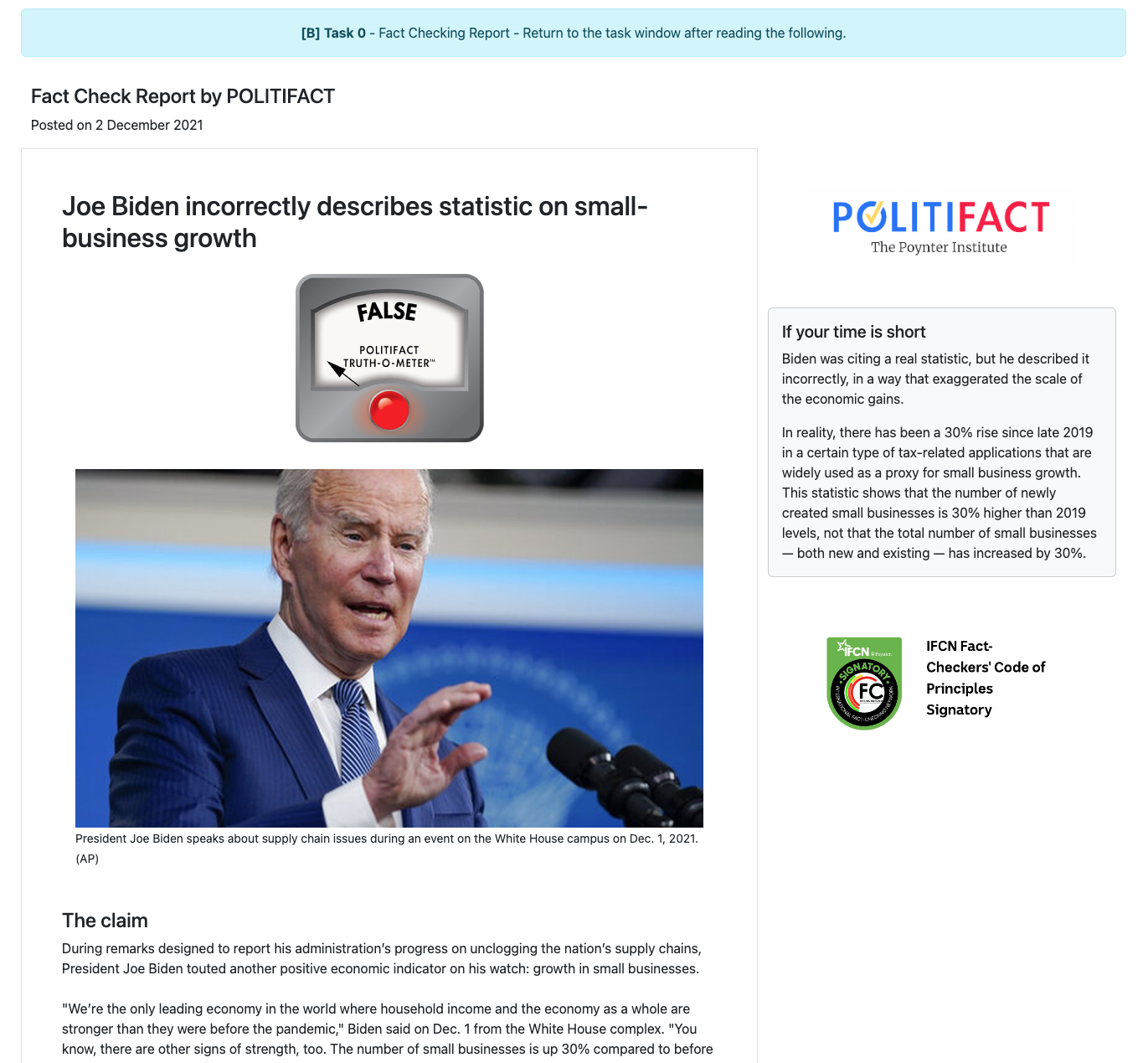}
         \caption{Improved Presentation and Credibility (IPC)}
         \label{fig:y equals x}
     \end{subfigure}
     \hfill
     \begin{subfigure}[b]{0.37\textwidth}
         \centering
         \includegraphics[width=\textwidth]{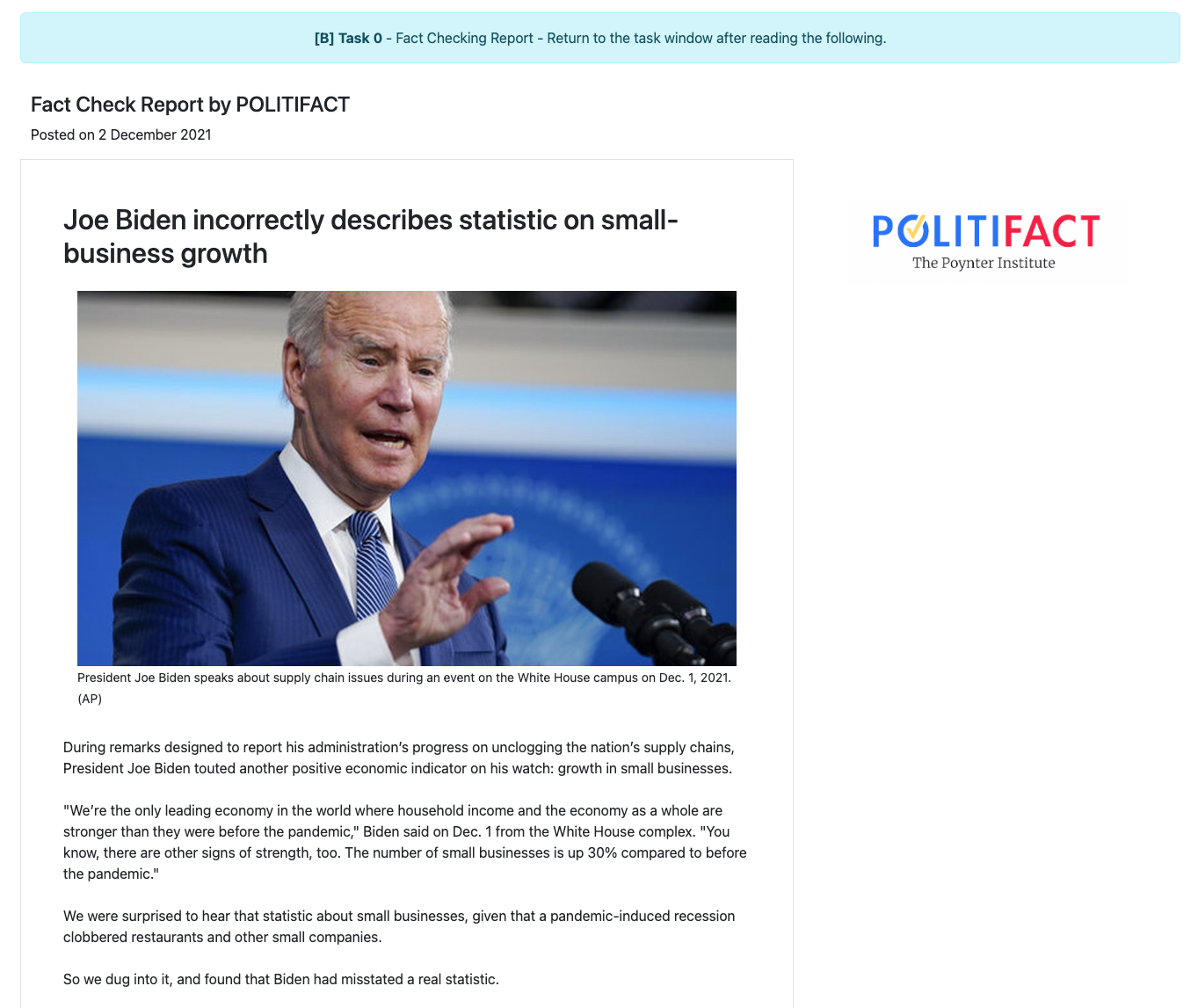}
         \caption{Baseline (B)}
         \label{fig:three sin x}
     \end{subfigure}
     \hfill
        \caption{Screenshots of fact-checking reports presented under different experimental conditions.}
        \label{fig:screenshots-reports}
\end{figure}

Based on the credibility and presentation features detailed in Table~\ref{tab:experiment-features}, we developed four study conditions: \emph{Baseline (B)}, \emph{Improved Presentation (IP)}, \emph{Improved Credibility (IC)}, and \emph{Improved Presentation and Credibility (IPC)}. As shown in Figure~\ref{fig:screenshots-reports}, our experiment setup allowed us to dynamically adjust the presentation of articles by adding necessary components according to the study condition. We used a within-subject experimental design, where all participants read and rated all eight fact-checking reports generated using the four experimental conditions. Each article was randomly assigned to one of the conditions such that each participant completed two articles (one from PolitiFact and one from the ABC) under each condition. Articles were also presented in randomized order.  

As shown in Figure~\ref{fig:study-flow}, each participant who launched the task from the Amazon Mechanical Turk platform completed three stages. In stage 1, participants completed two questionnaires. The five-item Conspiracy Mentality Questionnaire (CMQ)~\cite{Bruder2013-pf}, which measures susceptibility to conspiracy through statements such as `There are secret organizations that greatly influence political decisions'. 
The six-item Credibility of Science Scale (CoSS)~\cite{Hartman2017-ig} measures how individuals perceive scientific information and research as trustworthy, reliable, and valid. It includes statements such as `A lot of scientific theories are dead wrong.' and items are reverse scored with higher scores indicating a higher belief in science. For both measures, responses were recorded with a seven-point Likert scale ranging from `Disagree very strongly' to `Agree very strongly'. 

Stage 2 included eight tasks where users were asked to read a fact-checking report and answer a set of questions. In each task, the main task page included a link that opened the fact-checking report in a new browser tab. All user engagements (\eg~scrolls, clicks) within the report page were recorded and tracked. The questions included, (Q1) a binary response question asking for the reported veracity (true or false), (Q2) a multiple-choice question based on the content of the fact-checking report, (Q3) a free-text question asking users to explain in their own words how the fact-checking report verified the claim and (Q4) three rating questions that asked users whether the report (i) includes information that they would normally look for (i.e., completeness), (ii) appears to be credible (i.e., credibility) and (iii) is easy to follow (i.e., usability). Rating options were on a six-point Likert scale ranging from `disagree strongly' to 'agree strongly'. Finally, in Stage 3, participants completed a demographics survey that asked about their age, gender, highest level of education, political inclination, and general questions on experience with fact-checking.

The experiment was developed as a standalone web application using the Python Django framework and deployed on an online server.
The study was approved by RMIT University’s Ethics Board (ID 25436). We recruited participants through the online crowdsourcing platform Amazon Mechanical Turk. Participants were required to be located in the US, and have completed more than 1,000 tasks in the platform with an overall approval rate higher than 95\%. Each participant received US\$5 as the reward. The reward was calculated based on the average task completion time of our pilot testing and the minimum wage in the US. Furthermore, multiple-choice questions in the main task were used as quality checks.

\section{Results}

We recruited participants located in the US -- the recruitment from other English-speaking countries, including Australia, is planned for future work -- and used two quality control measures to select the final sample. First, we removed any participant ($n=10$) who did not view the report tab in at least one task (\ie~having no log events for the specific task). Second, based on the quality check question, we discarded responses from any participant who incorrectly answered more than four quality check questions ($n=13$). A total of 23 responses were discarded, resulting in responses from 76 participants considered for the evaluation. On average, participants spent 51 ($SD = 23$) minutes completing the entire study.

\subsection{User Interactions and Perceptions}

\subsubsection{Measured Outcomes}
We investigate the impact of presentation and credibility improvements on participants' ability to accurately interpret the verdict of the fact-checking articles, which we term as accuracy in our analysis. We calculated the binary accuracy for each article examined by each participant.

In our within-subject experiment, each participant rated articles related to all four study conditions. A non-parametric paired Wilcoxon signed rank test ($Z=148.5$, $p<0.05$) indicated a significantly higher task accuracy in conditions with presentation improvements (IP \& IPC, $M=89.1$) compared to conditions without presentation improvements (B \& IC, $M=82.9$). Considering credibility improvements, a Wilcoxon signed rank test ($Z=104.0$, $p<0.05$) indicated a significantly higher task accuracy in conditions with credibility improvements (IC \& IPC, $M=89.1$) compared to conditions without credibility improvements (B \& IP, $M=82.9$).

When considering other measured outcomes, we did not observe any significant differences resulting from either presentation or credibility improvements. However, presentation improvements (IP \& IPC) resulted in lower mean values in measures such as click count, scroll events, and mouse movements compared to conditions with no presentation improvements (B \& IC). Such lower values suggest that users could find relevant information and navigate the report quicker, with reduced interactions, and potentially a lower effort. In addition, when considering credibility improvements, mean values of measured outcomes -- such as task time, scroll events, and mouse movements -- were higher in conditions that included credibility improvements (IC \& IPC) compared to conditions that did not have such improvements (B \& IP). This observation indicates that additional elements introduced for improved credibility could potentially lead to information overload and reduced ease of access to critical information.

\begin{table}[tb]
\caption{Mean values of measured outcomes and user self-reported scores of the task based on \textit{Presentation}.}
\label{tab:metrics-pres}
\begin{minipage}{\columnwidth}
\begin{center}
\begin{tabular}{llSS}
\toprule
\multicolumn{2}{l}{\it Study Conditions} & {\it B} \& {\it IC}  & {\it IP} \& {\it IPC} \\
\multicolumn{2}{l}{Presentation Improvements} & {No}  & {Yes} \\
\midrule
{\it Measured} & Task accuracy * & 82.9 & 89.1 \\
{\it Outcomes} & Task time & 295.3 & 304.3 \\       
& Click count & 2.4 &   2.2 \\      
& Scroll events & 651.0 & 573.8 \\
& Maximum scroll depth & 4123.7 & 4144.5 \\         
& Mouse movements & 49.7 &  46.8 \\
\midrule
{\it Self-Reported} & Credibility & 4.94 & 4.90 \\            
{\it Scores} & Usability & 4.73 & 4.79 \\     
& Completeness & 4.38 & 4.33 \\
\bottomrule
\end{tabular}
\end{center}
\smallskip
\small * indicates a significant difference ($p<0.05$).
\end{minipage}
\end{table}

\begin{table}[tb]
\caption{Mean values of measured outcomes and user self-reported scores of the task based on \textit{Credibility}.}
\label{tab:metrics-cred}
\begin{minipage}{\columnwidth}
\begin{center}
\begin{tabular}{llSS}
\toprule
\multicolumn{2}{l}{\it Study Conditions} & {\it B} \& {\it IP}  & {\it IC} \& {\it IPC} \\
\multicolumn{2}{l}{Credibility Improvements} & {No}  & {Yes} \\
\midrule
{\it Measured} &Task accuracy * & 82.9 &     89.1 \\
{\it Outcomes} & Task time & 292.1 &    307.5 \\
& Click count & 2.5 &      2.1 \\
& Scroll events & 585.3 &    639.5 \\
& Maximum scroll depth & 3993.2 &   4275.1 \\
& Mouse movements & 47.0 &     49.6 \\
\midrule
{\it Self-Reported} & Credibility & 4.96 &     4.89 \\     
{\it Scores} & Usability & 4.81 &     4.71 \\
& Completeness & 4.40 &     4.31 \\
\bottomrule
\end{tabular}
\end{center}
\smallskip
\small * indicates a significant difference ($p<0.05$).
\end{minipage}
\end{table}

\subsubsection{Self-reported Scores}

In each task, using three representative questions, we asked participants to self-report how they rate the credibility, usability, and completeness of the article, as shown in Tables~\ref{tab:metrics-pres} and \ref{tab:metrics-cred}. We did not observe any statistically significant differences in Credibility, Usability, and Completeness. 

\subsection{User Attributes}

\begin{table*}[tb]
  \caption{Outcomes from the generalized linear model to predict  participants' ability to accurately interpret the
verdict of the fact-checking articles. * and boldface indicate significant predictors ($p<0.05$).}
  \label{tab:glm}
\adjustbox{max width=0.8\textwidth}{%
\begin{tabular}{p{10cm}rrrrrrrr}
\toprule
{\it Predictor} & {\it Coefficient}  & {\it Std.~Error} & {\it Z} & {\it $p$-value}  \\
\midrule
Gender -- Male  & -7.9614  & 4.064 & -1.959 & 0.050 \\
Age *             & 4.9810   & 2.251 & 2.213  & \textbf{0.027} \\
Highest level of education -- Diploma or equivalent        & 7.5143   & 6.677 & 1.125  & 0.260 \\
Highest level of education -- Secondary Education        & -5.3665  & 7.005 & -0.766 & 0.444 \\
Visiting fact-checking websites -- Occasionally (several times a month)       & -3.5781  & 5.662 & -0.632 & 0.527 \\
Visiting fact-checking websites -- Rarely (several times a year or less)            & 7.0069   & 7.386 & 0.949  & 0.343 \\
Self-reported ability to identify misinformation -- Moderately capable         & 4.9572   & 5.448 & 0.910  & 0.363 \\
Self-reported ability to identify misinformation -- A little bit capable      & 5.0823   & 7.189 & 0.707  & 0.480 \\
Political Alignment -- Independent & -5.1052  & 6.863 & -0.744 & 0.457 \\
Political Alignment -- Republican & 5.0222   & 4.927 & 1.019  & 0.308 \\
Previously shared misinformation -- debunked by a fact-checking article                  & 0.3632   & 4.834 & 0.075  & 0.940 \\
Previously shared misinformation -- debunked by another regular news article                & -7.7353  & 6.074 & -1.274 & 0.203 \\
Previously shared misinformation -- debunked by a social media user *               & -18.1271 & 8.674 & -2.090 & \textbf{0.037} \\
Previously shared misinformation -- debunked by someone they know             & -4.9829  & 6.571 & -0.758 & 0.448 \\
Credibility of Science Scale (CoSS) * & 3.7031   & 1.736 & 2.133  & \textbf{0.033} \\
Conspiracy Mentality Questionnaire (CMQ) & 1.2555   & 2.184 & 0.575  & 0.565 \\
\bottomrule
\end{tabular}
}
\end{table*}

Through pre- and post-task questionnaires, we collected participant demographics, and their responses to the Conspiracy Mentality Questionnaire (CMQ) and the Credibility of Science Scale (CoSS). Our study included 32 female, 44 male and 0 non-binary participants in age categories 18 to 24 ($n=1$), 25 to 34 ($n=22$), 35 to 44 ($n=36$), 45 to 54 ($n=12$), 55 to 64 ($n=4$), and 65 and above ($n=1$). 42 participants politically identified themselves as Democrats, 23 as Republicans, 10 as Independent and 1 as Other. Regarding the highest level of education, 54 had completed a Bachelor's Degree or higher, 10 had completed a Diploma or equivalent, 11 had completed Secondary Education (High School or Year 12), and 1 participant had completed Primary or Middle School.

When asked about how often they visit fact-checking websites, 20 stated `frequently (more than once a week)', 32 `occasionally (several times a month)' and 24 `rarely (several times a year or less)'. In terms of their self-reported capability at differentiating fake news from other kinds of news, 18 participants selected `very capable', 41 selected `moderately capable' and 15 selected `a little bit capable'. None of the participants indicated that they were not at all capable and 2 selected `unsure'. We also asked if they have ever shared a piece of information that they later discovered was false or misleading, 34 participants responded with a `no' while 42 said `yes'. From participants who previously shared misinformation, we asked how they discovered the shared information was false or misleading. 18 found such from a fact-checking article, 10 from another regular news article, 5 from a social media user, and 9 from someone that they know.

Using a generalized linear model, we analyzed the influence of the user attributes on their ability to accurately interpret the verdict of the fact-checking articles. As responses for the highest level of education completed did not have a reasonable response distribution across categories, we aggregated the responses into three categories (Bachelor's Degree or higher, Diploma or equivalent, Secondary Education (High School or Year 12)). We also represented age categories as ordinal values. We checked for the existence of multicollinearity to ensure the model validity. Our predictors indicate a maximum variance influence of 3.55, which is below the commonly accepted threshold of five to detect multicollinearity~\cite{Hair2013-fx}. The model was implemented using the Python \texttt{statmodels} library, with a resulting pseudo-R-squared value of 0.44. 

As detailed in Table~\ref{tab:glm}, the model indicated three significant predictors. Participants' age was a significant predictor. The results indicate that the ability to accurately interpret the fact-checking report increase with age. When considering previous encounters with sharing misinformation, participants who mentioned that they have previously shared misinformation, which they later discovered as misinformation through a social media user, had significantly lower ability to accurately interpret fact-checking reports compared to others who did not previously share misinformation. In addition, increased belief in science (\ie~a higher score on the Credibility of Science Scale) resulted in more accurate veracity judgments. 

\begin{figure}[tb]
    \centering
    \includegraphics[width=.95\columnwidth]{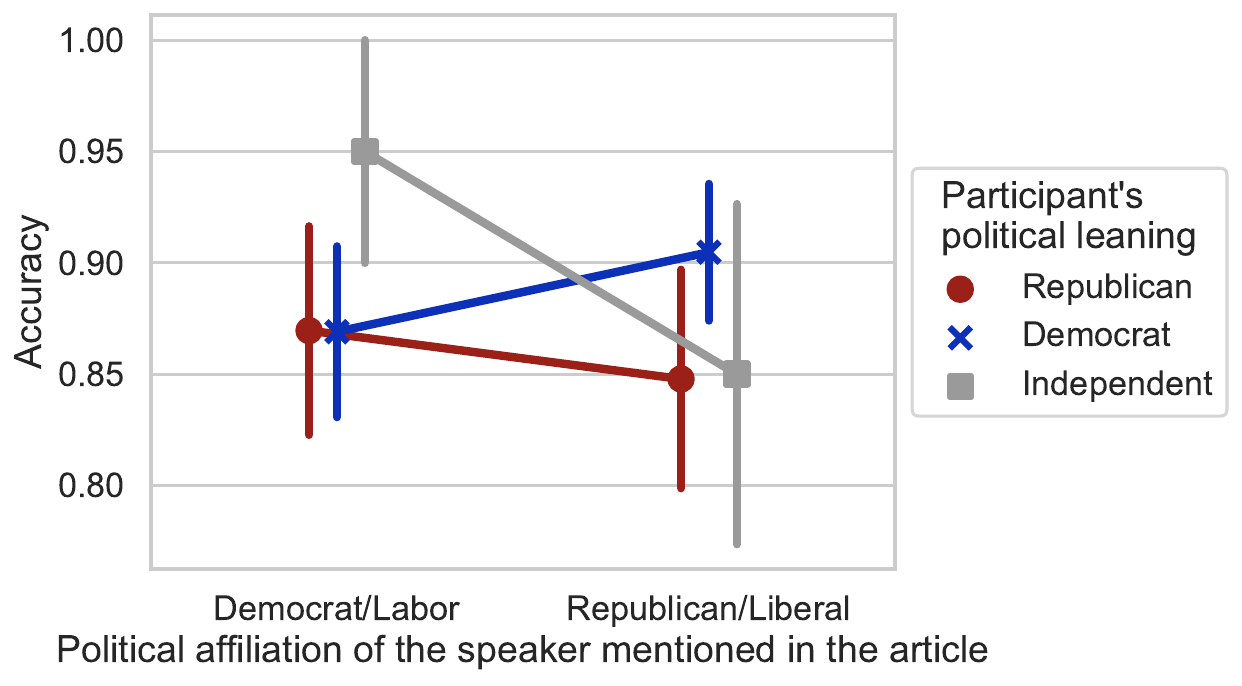}
    \caption{Task accuracy, participants' political leaning and the affiliation of the speaker described in the article. Error bars show standard error.}
    \label{fig:political}
\end{figure}

\subsubsection{Impact of Political Leaning}

Participants' political leaning alone did not appear as a significant predictor in our model (Table~\ref{tab:glm}). We further examined any potential interaction effects, as biases due to political leaning often appear as an important factor in the literature \cite{La_Barbera2020-qc,roitero2020can,epstein2020will}. As shown in Figure~\ref{fig:political}, participants tend to be better at accurately interpreting fact-checking reports that do not examine claims of politicians that align with their leaning. For example, Democrat-leaning participants have a higher task accuracy when rating fact-checking reports concerning claims by Republican/Liberal politicians compared to Democrat/Labor politicians.

Overall, to answer \textbf{RQ2}, we found that both credibility and presentation improvements significantly improve the users' ability to accurately interpret fact-checking reports. In addition, we show that user attributes such as age and belief in science can significantly impact the users' interpretation ability.

\section{Discussion}
We examine two aspects of our results and consider limitations.

\subsection{Designing Fact-Checking Reports}

Designing fact-checking reports requires careful consideration of the content, structure, and presentation of information. Based on our findings from the design workshop and online evaluations, we propose a set of design guidelines for effectively presenting fact-checking reports on screen-based interfaces. We organize these recommendations under presentation and credibility improvements and anticipate that these guidelines will help practitioners to create more effective fact-checking reports.

\textit{Presentation:} Communicating the final verdict in a straightforward manner will help a broader range of users, particularly those not interested in reading the entire article. Two basic steps are having a clear and consistent veracity indicator and phrasing the verdict through the report heading. Including a dedicated summary for each report is also helpful. Considering the main report's structure and text, it should be easy to read and understand. Use clear and concise language, and avoid technical jargon or complex sentence structures. Use headings, bullet points, and visual aids to break up the text and make it easier to digest. 

\textit{Credibility:} Ensure the report is credible by citing credible sources and providing links to additional information. It is essential to build trust with the user by providing details on who did the fact-checking and how. Information about accreditation and details about the author are simple additions that can potentially further enhance the perceived credibility of the report. However, as evident from our results, credibility additions can also be detrimental to the overall usability, increasing the user excise and the number of interactions required to find essential information in the report. 

\subsection{Individual Differences}

Our results on how user attributes influence the ability to accurately interpret a fact-checking report align with prior work on misinformation. Previous work by \acite{Saling2021-zz} reports significantly higher belief in science in the individuals who did not share possible misinformation than those who did. Similarly, our study reports that a higher belief in science leads to a better ability to accurately interpret fact-checking reports. These observations highlight the need for specific interventions and support for individuals with a lower belief in science.
Furthermore, prior work provides ample evidence on how users' political leaning bias their beliefs, judgments and actions related to misinformation~\cite{Pennycook2021-ka, La_Barbera2020-qc, Pereira2023-xz}. For instance, when crowd workers assess the truthfulness of statements, literature report that workers who voted for the speaker’s party provide higher scores for true statements~\cite{La_Barbera2020-qc}. As shown in Figure~\ref{fig:political}, we observe similar trends in how users interpret fact-checking reports. Participants tend to misinterpret the verdict when the fact-checking report discusses a claim from politically aligned speakers. 

While there is no disagreement around the effectiveness of fact-checking~\cite{Walter2020-ja,Nieminen2019-tw,Porter2021-sw,Carey2022-ob}, it has been noted that fact-checking reports can often fail to reach the target audience~\cite{Guess2020-ls}. As our workshop discussions on fact-checking consumption for various personas suggest, personalization could be an effective strategy to reach broader audiences. The need for personalization is further supported by our quantitative observations on how user attributes impact the ability to accurately interpret such reports.

\subsection{Limitations}

We note two limitations in our study. First, our task only included fact-checking reports that depicted a final verdict of either `true' or `false'. We adopted these binary categories to collect a total of 8 repeated judgments from each participant on a range of topics and multiple fact-checking outlets. Real-world fact-checking reports can contain verdicts on a broader truthfulness scale (\eg~Mostly False in PolitiFact) \cite{roitero2020can}, and future studies can adopt our experimental setup to investigate such scenarios. 
Second, due to the limited time and budget for each participant in our experimental setup, we opted to use three simple questions to capture self-reported user perception of the completeness, credibility, and usability of the reports. As evident from our results, potentially due to the anchoring bias of our Likert scale, we failed to collect conclusive evidence to investigate the impact on user perception. 
Future research can use existing validated instruments such as the system usability scale to better collect user perception. However, lengthy instruments may reduce worker engagement and lead to more task abandonment.

\section{Conclusion}

Through a collaborative workshop that involved fact-checking professionals, communications experts and researchers, we developed an understanding of the presentation of fact-checking reports for different interfaces. We then developed a crowdsourcing experiment to evaluate our presentation. Our results showed that both presentation and credibility improvements helped users to be more accurate when interpreting fact-checking reports. 
We anticipate that our findings and recommendations will be valuable for practitioners to refine their current report presentations and for further research that can improve how fact-checking reports are presented through screen and audio-based channels.
Building on our workshop findings, we plan to develop and evaluate presentation strategies for voice-based (\eg~voice assistants, podcasts) and other conversational (\eg~chatbots) interfaces in our future work.


\begin{acks}
This research is partially supported by the \grantsponsor{ARC}{Australian Research Council}{https://www.arc.gov.au/} (\grantnum{ARC}{CE200100005}, \grantnum{ARC}{DE200100064}, \grantnum{ARC}{DE200100540}). We thank Devi Mallal from RMIT FactLab and the participants of the workshop for their valuable contributions.
\end{acks}

\balance
\bibliographystyle{ACM-Reference-Format}
\bibliography{ref} 
\appendix

\end{document}